\documentclass[runningheads]{llncs}
\usepackage[T1]{fontenc}
\usepackage{graphicx}
\usepackage[dvipsnames]{xcolor}
\usepackage[utf8]{inputenc}
\usepackage[T1]{fontenc}
\usepackage{microtype}
\usepackage{listings}
\usepackage{makecell}
\usepackage{xargs}
\usepackage{lipsum}
\usepackage{xparse}
\usepackage{xifthen, xstring}
\usepackage{xspace}
\usepackage{marginnote}
\usepackage{etoolbox}
\usepackage[acronym,shortcuts]{glossaries}
\usepackage{amsmath}
\usepackage{thmtools} 
\usepackage{algorithm}
\usepackage{url}

\lstdefinelanguage{MLIR}{
  sensitive=true,
  morecomment=[l]{//},
  morecomment=[s]{/*}{*/},
  morestring=[b]",
  emph={builtin, func, device, hls, scf, arith}, 
  emphstyle=\color{gray},
  morekeywords={
    module, return,
    alloc, data_acquire, data_release,
    kernel_create, kernel_launch, kernel_wait,
    interface, axi_protocol, constant, index_cast,
    load, store, addf, subi, for, suite, group, scheme,
    table\_properties, arg\_table
  },
}

\lstdefinestyle{mlir_listing}{
  language=MLIR,
  basicstyle=\ttfamily\scriptsize,
  keywordstyle=\color{blue!60!cyan}\bfseries,
  emphstyle=\color{gray}\bfseries,
  commentstyle=\color{gray},
  stringstyle=\color{green!40!black},
  breaklines=true,
  showstringspaces=false,
  columns=flexible,
  numbers=none,
  frame=none,
  xleftmargin=0pt,
  framexleftmargin=0pt,
  escapeinside={(*@}{@*)} 
}

\lstdefinestyle{plain}{
  language      = {},               
  basicstyle    = \ttfamily\small,
  frame         = single,
  breaklines    = true,
  numbers       = left,
  numberstyle   = \tiny\color{gray},
}

\NewDocumentCommand{\mlirinline}{ m m }{%
  \textbf{\textcolor{gray}{#1}\textcolor{blue!60!cyan}{.#2}}%
}

\usepackage{color}

\begin{document}

\title{xDSL-CCPP: An alternative code generator for atmospheric modelling}
\author{Nick Brown\inst{1}\orcidID{0000-0003-2925-7275} \and
John Dennis\inst{2}\orcidID{0000-0002-2119-8242}}
\authorrunning{N. Brown and J. Dennis}
%
\institute{EPCC at the University of Edinburgh, Edinburgh, UK \and
National Center for Atmospheric Research (NCAR), Boulder, USA}
\maketitle              
\begin{abstract}
The Common Community Physics Package (CCPP) is a framework designed to simplify the exchange of physics packages between state-of-the-art atmospheric models.  A core software component of CCPP is a code generator: a bespoke Python-based application that uses regular expressions to transform an input specification into Fortran source code.  While the code generator initially provided a simple transformation capability, the demands of state-of-the-art atmospheric models have resulted in significant increases in its complexity. Furthermore, ongoing code modernisation efforts to enable GPU-based computing have placed additional demands on the code generator for which it was never designed.

Domain Specific Languages (DSLs) have proven to be highly effective in High Performance Computing (HPC), offering the ability to simplify code modernisation efforts. However, traditionally the effort and specialist skills required in the development of a DSL have been a blocker to adoption. In recent years, however, there have been great advances in compiler technologies, and frameworks such as MLIR and xDSL have the potential to greatly lower this barrier to entry.

In this paper, we present xDSL-CCPP, a new application-specific code generator based upon the latest MLIR and xDSL compiler technologies, for CCPP. Demonstrating a reduction in bespoke tooling code of around three times, from approximately 15,000 LoC to 5,000 LoC, we highlight the opportunities provided by these compiler frameworks and demonstrate how they significantly lower the barrier to development of HPC application-specific DSLs whilst maintaining application performance and promoting flexibility.

\keywords{MLIR, xDSL, Common Community Physics Package, Domain Specific Language (DSL)}
\end{abstract}
\section{Introduction}

The scientific computing community supports many large scientific code bases that must efficiently utilize modern High Performance Computing (HPC) computers.  Frequently, these large scientific code bases contain the contributions of a large number of domain scientists and represent 10 to 100 years of development effort.  The rapid advancement of HPC computing architectures typically means that these large scientific code bases are constantly being modernised.
In addition to the need to evolve for every change computer architecture, these codes are frequently focused on adapting how the end-user will interact with the reworked code. The Common Community Physics Package (CCPP) is one example of this, where the code has recently been modernised to simplify the exchange of physics packages between different climate and weather models.  Once a particular physics package has adopted the CCPP interface, it can be easily used by different CCPP-enabled models. However, flexibility of use comes at a cost. Specifically, it is based on a bespoke code generator (capgen) that automatically generates the Fortran interface code.  Furthermore, additional files that describe the transformation process and the source code must be maintained.

Effectively, the CCPP is a Domain Specific Language (DSL) in all but name. DSLs have proven to be highly effective; however, arguably, there are two major disadvantages to them that put people off. The first is the word \emph{language}, which implies that users must learn a new way of expressing their computation and/or configuration which is specific to the application in question. The second is the overhead in developing and maintaining bespoke compiler infrastructure to translate from the DSL to source code. Indeed, CCPP suffers from both of these issues, where configuration is undertaken by XML and metadata files that have grown unwieldy and the underlying Python-based tooling has become complex and difficult to maintain.

In this paper, we describe the development of xDSL-CCPP, which is a DSL for configuring the CCPP framework. Leveraging xDSL, a Python-based compiler framework that follows MLIR-based industry standard abstractions and concepts, we demonstrate that this not only significantly simplifies code generation tooling, but furthermore provides a range of additional opportunities, including embedding the DSL within an existing language that users are already familiar with. 

This paper is structured as follows: In Section \ref{sec:bg} we explore the background to this work, highlighting the role of CCPP and current code generation approaches along with the MLIR and xDSL compiler technologies. Section \ref{sec:xdsl-ccpp} then describes our xDSL-CCPP tool, which provides a compiler flow for CCPP code generation, along with exploring how this addresses limitations with the current bespoke approach. Section \ref{sec:results} then compares the code complexity, compile time, and resulting runtime of our approach against the existing approach. Lastly, Section \ref{sec:conc} draws conclusions and discusses further work.

\section{Background}
\label{sec:bg}

The Common Community Physics Package (CCPP) is a collaborative, open-source framework jointly developed by the National Center for Atmospheric Research (NCAR), the National Oceanic and Atmospheric Administration (NOAA), and additional collaborators via the Developmental Testbed Center (DTC). Modelling atmospheric physics within weather, climate, and Earth systems applications, CCPP is especially popular for exploring the integration, evaluation, and dissemination of physical parameterisations among research and operational entities \cite{bernardet2024common}.

There are two main components to CCPP: the first is the physics, which comprises the modelling of physical processes including radiation, convection, microphysics (water interaction), and land-surface interaction. Secondly, the framework serves as the foundational software architecture which connects different physics modules together and is the structure that enables users to leverage CCPP as a toolkit. The framework is designed to enable a \emph{pick and mix} approach to different configurations, where from one run to the next the set of enabled modules and their order can vary dramatically.

Several groups interact with CCPP, and we use the following terms in this paper. \emph{Scheme developers} are domain scientists who write and maintain individual physics schemes in Fortran, together with their metadata. \emph{Suite integrators}, whom we refer to as \emph{users}, assemble existing schemes into suites for a particular model configuration or experiment. These people write the suite configuration but do not modify scheme internals or the code generator. \emph{Host model developers} write the driver code of the atmospheric model that calls the generated interfaces. Lastly, \emph{tool maintainers} develop and maintain the code generator itself.

\begin{figure}[htb]
    \centering
    {\includegraphics[width=\columnwidth]{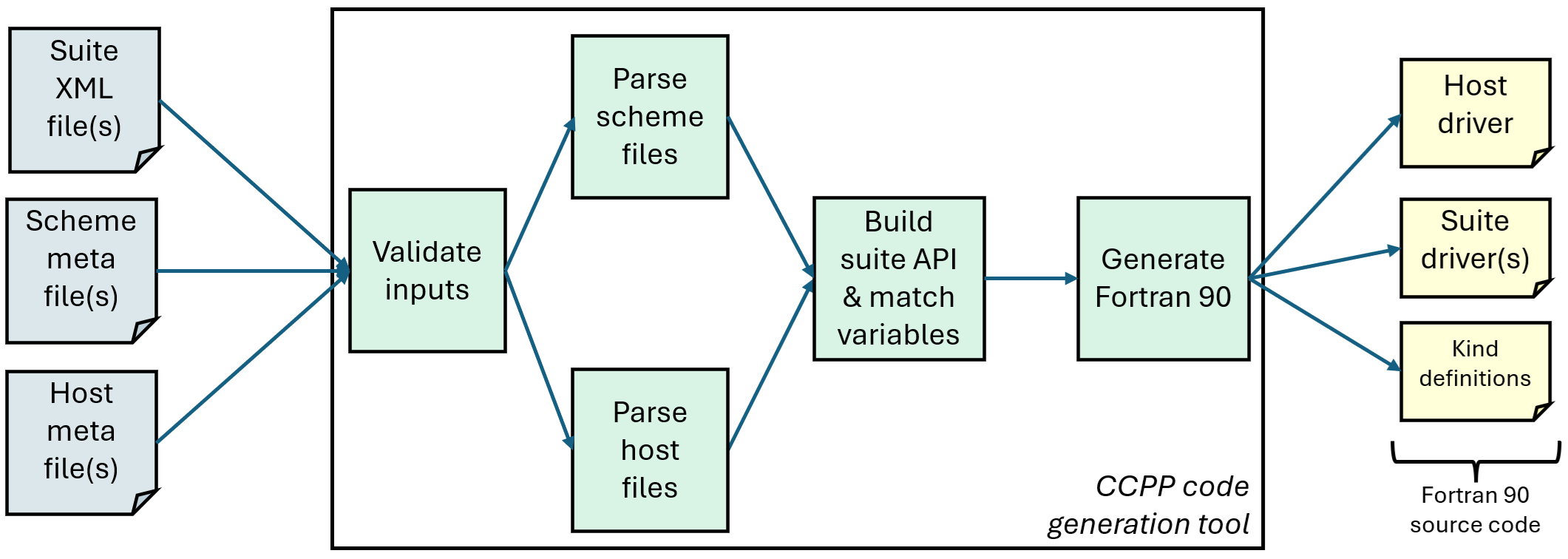}}
    \caption{Sketch of CCPP's existing code generation tool flow}
    \label{fig:ccpp-existing}
\end{figure}

CCPP is currently in its sixth version, and has recently been extensively rewritten to adopt modern Fortran standards. The code follows a modular approach, where each physical process is represented as an independent \emph{scheme}. One run to the next often comprises a different range of physical schemes run in different orders, and consequently a code generation approach based upon configuration files has been developed. Suite integrators provide an XML \emph{suite}, illustrated in Listing \ref{lst:suite}, along with metadata about the schemes and host (driver) Fortran code, sketched in Listing \ref{lst:metadata}. The tool flow is illustrated in Figure \ref{fig:ccpp-existing}, where XML and metadata files are provided as inputs to the CCPP code generation tool. This tool starts by validating and then parsing the input configuration files, before internally generating the required API calls and matching them to variables across schemes, with Fortran 90 source code files generated as output. This generation tool is written in Python, with the input configuration being a mixture of XML that describes the suite(s) of physics schemes to run, and key-value pair files for metadata that describe each global variable and procedure along with each of its arguments.

\begin{lstlisting}[style=plain, label=lst:suite, caption=Suite XML configuration, numbers=none, frame=none, xleftmargin=0pt, framexleftmargin=0pt]
<suite name="hello_world_suite" version="1.0">
  <group name="physics">
    <scheme>hello_scheme</scheme>
    <scheme>temp_adjust</scheme>
  </group>
</suite>
\end{lstlisting}

The \emph{hello\_world\_suite} of Listing \ref{lst:suite} is deliberately minimal for presentation and, in practice, a suite such as those used in Section~\ref{sec:results} orders tens of schemes. These cover radiation, boundary-layer turbulence, convection and cloud microphysics, and some schemes, typically microphysics, are sub-cycled several times within each physics timestep. Integrators frequently require variants of a suite that differ only in a single scheme or a tunable parameter, and the same scheme is reused across multiple host models. This is why scheme interfaces are described by host-independent metadata.

This code generation approach started as a fairly small Python script, but has grown significantly to support an increasing level of configuration complexity. There is a lot of boilerplate metadata for each Fortran source file, some of which can be mechanically generated while other components must be written by hand.  An example is provided in Listing \ref{lst:metadata}, which sketches the metadata of the \emph{hello\_scheme\_init} procedure, which is part of the \emph{hello\_scheme} scheme, and the file contains key-value pairs describing the signature of the procedure. Not only is such a manual metadata approach potentially error-prone, but furthermore the fixed nature of these files means that it is not currently possible to specialise them at configuration time, which results in duplicate schemes with only very minor differences. 

\begin{lstlisting}[style=plain, label=lst:metadata, caption=Extract of suite metadata for hello\_scheme's hello\_scheme\_init procedure, numbers=none, frame=none, xleftmargin=0pt, framexleftmargin=0pt]
[ccpp-table-properties]
  name = hello_scheme
  type = scheme
[ccpp-arg-table]
  name = hello_scheme_init
  type = scheme
[ errmsg ]
  standard_name = ccpp_error_message
  long_name = Error message for error handling in CCPP
  units = none
  dimensions = ()
  type = character
  kind = len=512
  intent = out
[ errflg ]
  standard_name = ccpp_error_code
  long_name = Error flag for error handling in CCPP
  units = 1
  dimensions = ()
  type = integer
  intent = out
\end{lstlisting}

The underlying issue is that the parsing and code generation framework has grown significantly in complexity to over 15000 lines of code and become brittle. There are numerous assumptions hardcoded into this current set of scripts, meaning that it is risky to add new functionality. Furthermore, it would be a significant amount of work to adopt an alternative user interface, for instance replacing the current XML and metadata with Python, although that is potentially desirable as it would provide more flexibility, especially around better representing certain constructs such as loops. 

\subsection{MLIR and xDSL}

LLVM \cite{lattner2004llvm} is a collection of reusable compiler and toolchain components for building compilers for many source languages and deploying to a wide variety of targets. It offers multiple language frontends and many architecture-specific backends, with LLVM-IR acting as the common interface between them. As a result, a frontend such as Clang (for C and C++) can emit LLVM-IR and immediately gain access to any supported backend, including those for CPUs, GPUs, and FPGAs. A drawback of this design is that LLVM-IR is relatively low-level, so frontends must implement substantial lowering logic themselves, often leading to duplicated effort across different frontends.

MLIR, originally developed at Google and later released as open source, addresses this by introducing a hierarchy of intermediate representations called dialects together with reusable transformations between them. Rather than requiring frontends to target LLVM-IR directly, MLIR allows them to emit higher-level IR and then apply shared lowering and optimisation passes to progressively reduce abstraction until reaching LLVM-level code generation. Intermediate Representation (IR) is manipulated through a series of transformation passes, lowered from one to the next, ultimately resulting in IR that can be consumed by a backend such as LLVM. IR follows a standard form known as Static Single Assignment (SSA), and a key feature is that multiple dialects can coexist within the same IR module and be transformed independently, supporting staged lowering toward the LLVM dialect and ultimately to LLVM-IR. This structure encourages reuse of compiler infrastructure across frontends, reducing the engineering cost of developing new compilers, while also providing extension mechanisms for defining custom dialects and passes.

Since becoming an LLVM sub-project, MLIR has gained significant adoption and an active development community. The project ships a broad set of standard dialects, including arith for arithmetic operations, scf for structured (serial and parallel) control flow, affine for affine loop representations, memref for memory and data access, func for functions and calls, openmp for OpenMP constructs, linalg for linear algebra operations, and vector for vectorisation. These dialects can be lowered to the llvm MLIR dialect, from which LLVM-IR is typically produced using the mlir-translate tool.

A practical difficulty with MLIR is its substantial learning and development overhead: effective use commonly requires C++ expertise, familiarity with LLVM’s design principles, use of TableGen \cite{hsu2021llvm} to specify dialect interfaces, and continuing maintenance effort in response to a rapidly evolving codebase.

In contrast, xDSL \cite{xdsl} is a Python-based compiler toolkit that is designed to be compatible with MLIR at a one-to-one level. It re-implements most standard MLIR dialects and also includes additional experimental dialects, with dialect definitions expressed using IRDL \cite{fehr2022irdl} as Python classes. xDSL is intended to support fast iteration and prototyping of MLIR-like ideas, which can later be upstreamed into MLIR once they mature; the MPI dialect provides an example, having been developed first in xDSL \cite{bisbas2024shared} before subsequent standardisation in MLIR. Because of this close compatibility, it is possible to move between xDSL and MLIR during a compilation workflow. In this work, we employ xDSL to implement the transformations and optimisations presented in the paper.

A major purpose of xDSL has been to lower the barrier to developing DSL compilers by promoting the sharing of infrastructure and avoiding siloing of the underlying compiler stack. This was demonstrated in \cite{bisbas2024shared} where the compilers of Devito and PSyclone DSLs were integrated into xDSL. Whilst these two DSLs are very different, the former is Python-based for seismology simulation and the latter Fortran-based used by the weather and climate community, there was around a 95\% reduction in compiler code size for PSyclone \cite{bisbas2024shared} resulting from this integration. There have been several other successes in using xDSL, for instance stencils running on the Cerebras \cite{stawinoga2026mlir} and an OpenMP flow for FPGAs \cite{rodriguez2025mlir} and AMD AIEs \cite{brown2026ccgrid}; however, until this work it has not yet been used as infrastructure for replacing existing configuration tooling and supporting the formalisation of configuration via DSLs. However, the key benefits around sharing of infrastructure to lower the overhead required for individual DSLs can also apply here.

\section{xDSL-CCPP: An xDSL-based code generation framework}
\label{sec:xdsl-ccpp}

\label{sec:our-dsl}
\begin{figure*}[htb]
    \centering
    {\includegraphics[width=\textwidth]{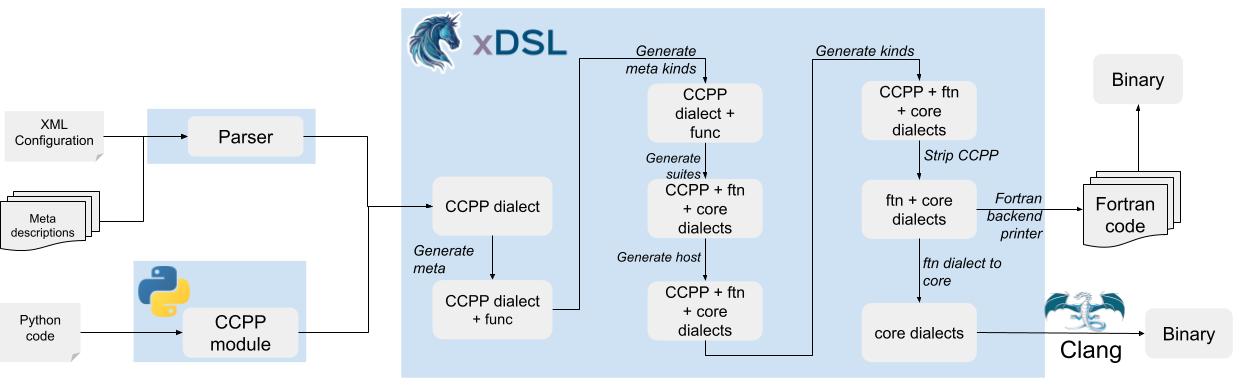}}
    \caption{Illustration of our xDSL-CCPP compiler flow, accepting either XML and metadata files or Python code, and generating either Fortran source code or LLVM-IR which can be provided to Clang}
    \label{fig:compiler_flow}
\end{figure*}

Figure \ref{fig:compiler_flow} illustrates the compiler flow of our approach, which leverages xDSL. We developed a bespoke dialect, \emph{ccpp}, which represents the Intermediate Representation (IR) level description of suite(s) and metadata. The user's XML configuration and metadata files are consumed by a Python-based frontend which parses the XML using the standard Python library and the key-value pairs in the metadata to generate IR containing this \emph{ccpp} dialect. This is illustrated in Listing \ref{lst:ccpp_mlir} where \mlirinline{ccpp}{suite} defines the suite, \emph{hello\_world\_suite}, which contains two schemes \emph{hello\_scheme} and \emph{temp\_adjust}. The \mlirinline{ccpp}{table\_properties} operation contains metadata for each scheme, \emph{hello\_scheme} in Listing \ref{lst:ccpp_mlir}. This comprises the signature of each procedure, and Listing \ref{lst:ccpp_mlir} includes the definition of the \emph{hello\_scheme\_init} procedure, which has two arguments \emph{errmsg} and \emph{errflg}.

\begin{lstlisting}[style=mlir_listing, label=lst:ccpp_mlir, caption=Sketch of IR with CCPP suite and meta description]
"ccpp.suite"() <{suite_name = "hello_world_suite", version = "1.0"}> ({
    "ccpp.group"() <{group_name = "physics"}> ({
      "ccpp.scheme"() <{scheme_name = "hello_scheme"}> : () -> ()
      "ccpp.scheme"() <{scheme_name = "temp_adjust"}> : () -> ()
    }) : () -> ()
}) : () -> ()

"ccpp.table_properties"() <{name = "hello_scheme", type = #ccpp<table_type_kind scheme>}> ({
    "ccpp.arg_table"() <{name = "hello_scheme_init", type = #ccpp<table_type_kind scheme>}> ({
      "ccpp.arg"() <{name = "errmsg", type = "character", standard_name = "ccpp_error_message", long_name = "Error message for error handling in CCPP", kind = "len=512", intent = "out", units = "none"}> : () -> ()
      "ccpp.arg"() <{name = "errflg", type = "integer", standard_name = "ccpp_error_code", long_name = "Error flag for error handling in CCPP", intent = "out", units = "1"}> : () -> ()
    }) : () -> ()
}) : () -> ()
\end{lstlisting}

The IR generated by the frontend is then consumed by our xDSL compiler pipeline where it is manipulated via a series of transformations developed in this work. The first of these transformations, \emph{generate\_meta}, interrogates the metadata within the IR and generates corresponding function signature definitions. This is required because subsequent pipeline stages insert operations that issue calls to these functions, and so the IR must contain the signature of these functions for error checking. 

Whilst the Flang Fortran compiler is built upon MLIR, its own dialects \emph{fir} and \emph{hlfir} are extensive and fairly low-level, not necessarily representing higher-level Fortran constructs. Consequently, we have developed our own \emph{ftn} dialect, which contains the Fortran-specific types and operations required which includes the definition of kinds, string comparison and string assignment. The next transformation, \emph{generate\_meta\_kinds} also operates on the \emph{ccpp} dialect metadata held within the IR, and identifies any required Fortran kinds. These are specifically included for real number representation, such as the type \emph{real(kind=k)}, and the corresponding kind attribute and definition operation from the \emph{ftn} dialect are inserted into the IR. 

At this point, the IR contains enough information upon which code-level IR constructs can be generated. Upstream MLIR contains a wide range of core dialects that describe code level aspects such as control flow (\emph{scf}), memory allocation and access (\emph{memref}) and functions (\emph{func}). Consequently, the next transformation, \emph{generate\_suites}, consumes the suite description held in the IR by the \emph{ccpp} dialect, and generates code level IR containing the core MLIR dialects. The constituent functions for suite initialisation, finalisation, and execution are generated, and Listing \ref{lst:ccpp_suite} sketches the \emph{hello\_world\_suite\_initialize} function that has been generated from the suite and metadata of Listing \ref{lst:ccpp_mlir}. This function accepts no arguments and produces two outputs which are memory references to data (an array). The \mlirinline{memref}{alloc} operations at the start of the function allocate these arrays, and the value of 0 is then written into \emph{errflg} to denote no initial error. 

It can be seen from Listing \ref{lst:ccpp_suite} that the generated initialisation function then calls into individual scheme initialisations if \emph{errflg} equals zero, with these scheme functions also returning an \emph{errflg} and \emph{errmsg} which are copied into this function's memrefs using the \mlirinline{memref}{copy} operation. Lastly, \emph{errflg} and \emph{errmsg} are returned via the \mlirinline{func}{return} operation to the caller. Other suite functions follow a similar structure, albeit more complex, to the initialisation sketched in Listing \ref{lst:ccpp_suite}. 

\begin{lstlisting}[style=mlir_listing, label=lst:ccpp_suite, caption=Sketch of generated \emph{hello\_world\_suite\_initialize} subroutine which calls individual scheme initialisation]
func.func public @hello_world_suite_initialize() -> (memref<512xi8>, memref<i32>) {
    %errmsg = "memref.alloc"() <{operandSegmentSizes = array<i32: 0, 0>}> : () -> memref<512xi8>
    %errflg = "memref.alloc"() <{operandSegmentSizes = array<i32: 0, 0>}> : () -> memref<i32>

    %0 = arith.constant 0 : i32
    memref.store %0, %errflg[] : memref<i32>

    %10 = arith.constant 0 : i32
    %11 = memref.load %errflg[] : memref<i32>
    %12 = arith.cmpi eq, %11, %10 : i32
    scf.if %12 {
        %13, %14 = func.call @hello_scheme_init() : () -> (memref<512xi8>, memref<i32>)
        "memref.copy"(%13, %errmsg) : (memref<512xi8>, memref<512xi8>) -> ()
        "memref.copy"(%14, %errflg) : (memref<i32>, memref<i32>) -> ()
    }
      
    %15 = arith.constant 0 : i32
    %16 = memref.load %errflg[] : memref<i32>
    %17 = arith.cmpi eq, %16, %15 : i32
    scf.if %17 {
        %18, %19 = func.call @temp_adjust_init() : () -> (memref<512xi8>, memref<i32>)
        "memref.copy"(%18, %errmsg) : (memref<512xi8>, memref<512xi8>) -> ()
        "memref.copy"(%19, %errflg) : (memref<i32>, memref<i32>) -> ()
    }
    func.return %errmsg, %errflg : memref<512xi8>, memref<i32>
}
\end{lstlisting}

The next transformation, \emph{generate\_host}, generates IR that will be called by the overarching driver code for the entire application initialisation, finalisation and execution. This code calls into the corresponding suite functions as required, and marshals control between them. The \emph{strip\_ccpp} transformation now removes CCPP-specific configuration and metadata from the IR, with the IR now only containing code-level operations which closely represent the execution level.

At this point we have one of two options in the compilation flow: either generating Fortran code directly from the IR and providing this as an input to any compiler, or passing the IR directly into LLVM via Clang and compiling to object files. The first of these is how the current framework operates and as such we developed a Fortran backend which consumes the IR and generates Fortran 90-conforming code based upon this. A mapping is provided between each operation, such as the subroutine definition based upon the \mlirinline{func}{func} (function definition) operation or a loop construct from \mlirinline{scf}{for}. Expressions are also mapped to their Fortran equivalents, which tends to be a straightforward conversion. Fortran-specific constructs are represented by operations in our bespoke \emph{ftn} dialect, such as string equality and kind definition, and this simplifies code generation as there is a one-to-one mapping here.

Alternatively, the IR can be provided to LLVM via Clang to produce an object file which is then linked against the rest of the code. Whilst this avoids the need to generate Fortran code and leverage a separate compiler, it does tie the entire flow into LLVM, as object files from other parts of the code must be compatible with the object files generated here. The \emph{ftn\_dialect\_to\_core} transformation lowers constructs in our \emph{ftn} dialect into core MLIR/LLVM dialects so that the IR can be consumed by LLVM.

\subsection{Removing the need for manual metadata files}
\label{sec:F90meta}
One of the major limitations with the current CCPP code generation approach is that scheme developers must manually maintain metadata key-value pair files similar to that sketched in Listing \ref{lst:metadata}. While CCPP provides the capability to generate several of these fields mechanically, additional fields must be added and maintained by hand. Not only is this time consuming and error-prone, but it also imposes a maintenance burden where metadata and source files must remain up to date with respect to each other. As xDSL is one-to-one compatible with MLIR, it is possible to leverage the large ecosystem that has grown up around MLIR to help address this. In 2020, the Flang Fortran compiler \cite{flang} was rewritten from the ground up to support modern Fortran standards and built upon MLIR. To this end, we can leverage Flang as part of our compiler pipeline to automatically generate this metadata and avoid manually written metadata files. We developed a new transformation, \emph{fir\_to\_meta}, which consumes the IR generated by Flang and deduces the metadata based upon the details encoded within.  

Even though Fortran sits outside the core MLIR ecosystem, xDSL provides both \emph{hlfir} and \emph{fir} dialects, so our transformation, \emph{fir\_to\_meta}, interrogates each IR generated by Flang from Fortran source code and from this generates the corresponding CCPP metadata IR. Consequently, by leveraging xDSL and MLIR, now only the suite XML configuration file(s) need to be provided to the configuration flow, as metadata can be automatically inferred by leveraging the wider MLIR ecosystem.

\subsection{A Python-based DSL}

As illustrated in Figure \ref{fig:compiler_flow} the CCPP dialect, generated by the XML Python frontend, is the front door into our compiler pipeline, which then contains the majority of the functionality. This thin transformation layer from the user's interface to the CCPP IR currently operates upon XML and metadata files, the latter either written explicitly or inferred from the Fortran source code. However, the IR can be trivially generated from any frontend as long as it follows the expected format. This is illustrated by the Python frontend in Figure \ref{fig:compiler_flow} and enables us to readily drive the configuration by alternative user interfaces.

This addresses an important problem because a challenge currently faced is that suite integrators need to represent more complex structures in their suite configuration, such as loops, which do not necessarily map naturally to XML. Consequently, we developed a Python-based configuration frontend where scientists describe their flow in Python code, and the imported Python module interrogates this and generates the corresponding CCPP IR that is compatible with the rest of the pipeline.

\begin{lstlisting}[language=Python, label=lst:pyfront, caption=Sketch of suite definition using Python frontend, frame=none, xleftmargin=0pt, framexleftmargin=0pt]
from xdsl_ccpp.frontend.py_api import Arg, ccpp_scheme, ccpp_suite

@ccpp_suite("hello_world_suite", version="1.0")
class hello_world:
    physics = [hello_scheme, temp_adjust]

    def run():
        physics()

@ccpp_scheme
class hello_scheme:
  init = [errmsg, errflg]
\end{lstlisting}

The Python representation of suite and scheme definitions from Listing \ref{lst:suite} are illustrated in Listing \ref{lst:pyfront}. This frontend leverages the metadata inference approach described in Section \ref{sec:F90meta}, avoiding the need for this information to be provided by the programmer. Lines 3 to 8 in Listing \ref{lst:pyfront} define the suite, and lines 10 to 12 the scheme. The suite definition of lines 3 to 8 groups each constituent scheme, in this case into the \emph{physics} group, and the default is for each initialisation, execution and finalisation phase to call each group, and the schemes within a group, in the order that they are declared. However, it is also possible for the programmer to provide optional \emph{init}, \emph{run} and \emph{finalise} methods that override this default. For instance, in the \emph{run} method at lines 7 and 8 of Listing \ref{lst:pyfront}, and this is useful to support additional constructs and specialisation such as conditionals around scheme execution and loops.

Python enables control flow to be represented more naturally than via XML files and also enables scheme parameterisation, where physics schemes are specialised at code generation time that currently requires duplicate schemes with the XML approach. This is illustrated in Listing \ref{lst:pytwo}, where the \emph{run} function iterates over both schemes ten times and provides the loop counter to the scheme as well as all the other arguments. Furthermore, the arguments \emph{init\_val=98, diagnostics=False} to \emph{hello\_scheme} illustrate parameterisation of this scheme, where these arguments are set at compile time and enable convenient specialisation. 

\begin{lstlisting}[language=Python, label=lst:pytwo, caption=Sketch of more advanced suite definition in Python supporting loops and parameterisation of suites at compile time, frame=none, xleftmargin=0pt, framexleftmargin=0pt]
from xdsl_ccpp.frontend.py_api import ccpp_suite

number_of_subcycles=10

@ccpp_suite("hello_world_suite", version="1.0")
class hello_world:
    physics = [hello_scheme, temp_adjust]

    def run():
        for i in range(0, number_of_subcycles):
            hello_scheme(i, init_val=98, diagnostics=False)
            temp_adjust(i)
\end{lstlisting}

One of the major advantages of the automation of metadata discovery and the Python frontend is that this helps address the \emph{language} criticism of DSLs. Instead, we demonstrate that by leveraging xDSL and MLIR, one is able to significantly simplify the interface, in this case removing the need for manually written metadata files and embedding the suite configuration within Python that the users are already familiar with.

\section{Results and evaluation}
\label{sec:results}
Table \ref{tab:loc} reports a Lines of Code (LoC) comparison (ignoring code comments) between the existing CCPP code generation tool and our approach, xDSL-CCPP. Whilst lines of code is not an exact measure of code complexity, it still provides a measure of potential maintenance burden. It can be seen that our approach requires around three times fewer lines of code because it leverages xDSL, and indeed our xDSL-CCPP also includes the metadata discovery and Python-based frontend, which the existing approach does not provide.

\begin{table}[htb]
    \centering
    \caption{Lines of code comparison (ignoring code comments) between the existing CCPP code generation tool and our xDSL-CCPP approach}
    \label{tab:loc}
    \begin{tabular}{|c|c|}
    \hline     
      \textbf{Approach} & \textbf{Lines of code} \\
      \hline
    Existing & 15054 \\
    xDSL-CCPP & 4859 \\    
    \hline
    \end{tabular}
\end{table}

We now leverage three test cases to compare the existing configuration approach against our xDSL-CCPP framework.  The \emph{advection} test case simulates atmospheric advection, which is the movement of particles through the air, for 100,000 columns; the \emph{ddthost} test case models simplified atmospheric physics parameterisations, providing a self-contained test of how host-provided state in Derived Data Types (DDT) drives physics computations, such as temperature-related adjustments, across 100,000 columns. Lastly, \emph{capgen} focuses on thermodynamic and atmospheric state adjustment, also across 100,000 columns. In these test cases, the physics serves as a vehicle to exercise and test the underlying code generation and model integration features, such as how DDTs with temperature fields are represented, how dimensions affect allocations across slices of data, and the interaction between multiple suites and schemes.

Averaged over three runs, Table \ref{tab:compile_time} reports compile time for each of the three test cases on Cirrus, a Cray EX4000, comparing the existing code generation tool to xDSL-CCPP. It can be seen that our approach is slower than the existing tool, and the majority of this time is within the compilation pipeline, with the overhead being the generation of IR that is then marshalled between different transformations. However, the code generation time when using xDSL-CCPP is still less than a second.

\begin{table}[htb]
    \centering
    \caption{Code generation time comparison between the existing tool and our xDSL-CCPP}
    \label{tab:compile_time}
    \begin{tabular}{|c|c|c|}
    \hline     
      \textbf{Test case} & \makecell{\textbf{Existing} \\ \textbf{compile time (s)}} & \makecell{\textbf{xDSL-CCPP} \\ \textbf{compile time (s)}} \\
      \hline
    advection & 0.27 & 0.79 \\
    capgen & 0.29 & 0.86 \\    
    ddthost & 0.28 & 0.84 \\    
    \hline
    \end{tabular}
\end{table}

\begin{figure}[htb]
    \centering
    {\includegraphics[width=\columnwidth]{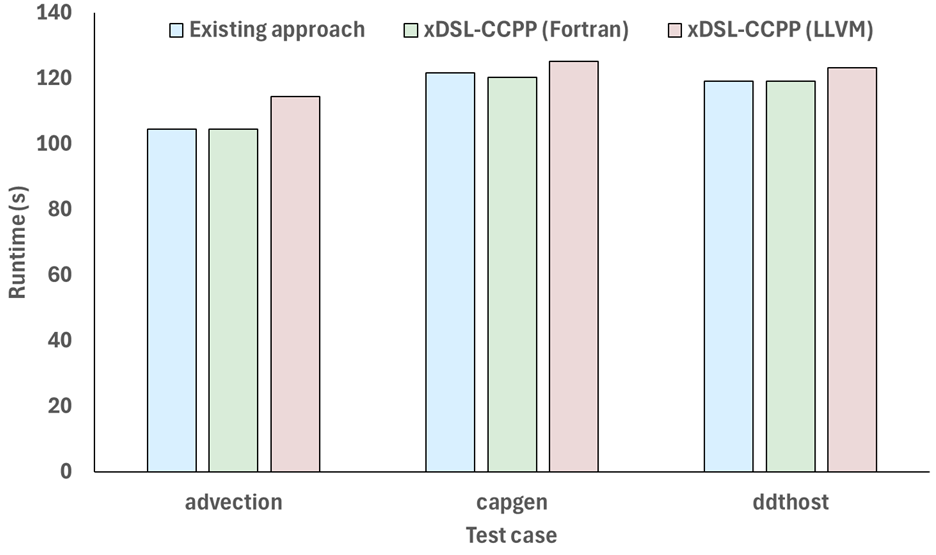}}
    \caption{Single core runtime performance comparison across three test cases on Cirrus, which is an HPE Cray EX4000}
    \label{fig:ccpp-cirrus}
\end{figure}

Figure \ref{fig:ccpp-cirrus} reports single-core runtime performance of the three test cases on Cirrus, a CPU-based HPE Cray EX4000 with AMD Zen5 (Turin). \emph{Existing approach} reports performance of the existing CCPP code generation tool and it can be seen that this is comparable to our \emph{xDSL-CCPP (Fortran)} approach that is using the Fortran code generator flow and then compiled along with the rest of the Fortran source code. These two tests leverage GCC version 14.2, and as described in Section \ref{sec:our-dsl} an alternative approach is to instead lower directly to LLVM-IR which is then built by Clang, and Flang compiles the scheme Fortran source code. LLVM version 20.1.7 is used here, and it can be seen in Figure \ref{fig:ccpp-cirrus} that this is slightly slower than GCC v14.2. Consequently, the Fortran backend is the generally preferred approach for xDSL-CCPP, but having an LLVM-IR route provides additional future flexibility.

\begin{figure}[htb]
    \centering
    {\includegraphics[width=\columnwidth]{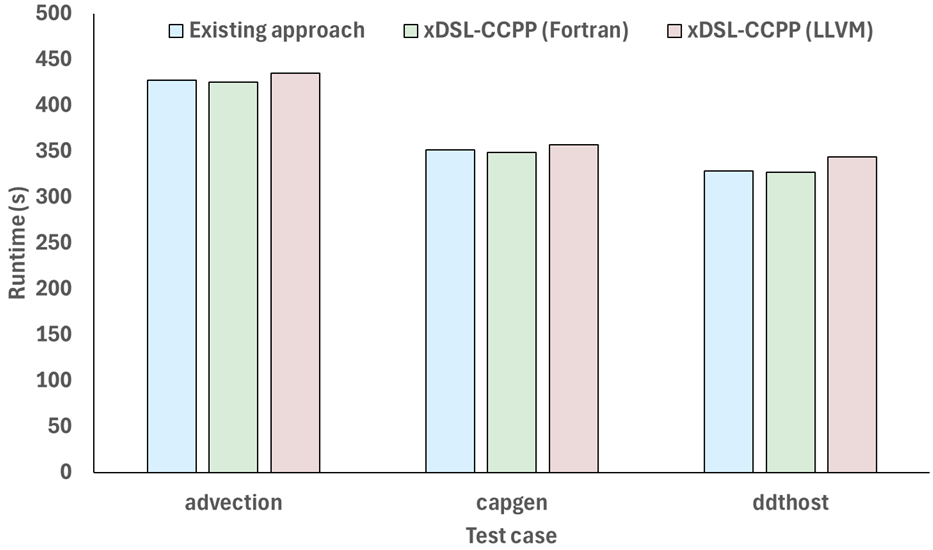}}
    \caption{Single core runtime performance comparison across three test cases on ARCHER2 which is an HPE Cray EX}
    \label{fig:ccpp-archer2}
\end{figure}

Figure \ref{fig:ccpp-archer2} reports the same performance comparison as in Figure \ref{fig:ccpp-cirrus}, but this time on ARCHER2, which is a Cray EX and the UK national supercomputing service. ARCHER2 is a CPU-based HPC machine containing AMD Zen2 (Rome), whilst the relative performance between the approaches is unchanged, it can be observed that absolute runtime on ARCHER2 is much longer and this is due to Zen2 in ARCHER2 compared to Zen5 in Cirrus. Moreover, on ARCHER2 the \emph{advection} testcase is slower than the other two, whilst on Cirrus it is faster. This is because this test case is the most computationally intensive, and AMD Zen2 only provides AVX2 (256 bits wide) whereas Zen5 provides AVX512 (512 bits wide), the latter being important for compute performance. 

\subsection{Single Column Model (SCM)}

The test cases explored so far focus on correctness of realistic physics data flows and computations, rather than validating scientific accuracy. NCAR's Single Column Model (SCM) is built upon CCPP and we integrated xDSL-CCPP into the build flow. We have compared our approach against the existing code generator across four SCM suites on Cirrus, all compiled with GCC v14.2 and using the xDSL-CCPP Fortran backend flow.

The first suite is LASSO (Large eddy simulation ARM Symbiotic Simulation and Observation), modelling continental shallow convection based on input data from 2nd of October 2018. GABLS3 (GEWEX Atmospheric Boundary Layer Study) is used for the development of a nocturnal jet based on data from 1st of July 2006; BOMEX (Barbados Oceanographic and Meteorological EXperiment) simulates maritime shallow convection, and lastly COMBLE (Cold-Air Outbreaks in the Marine Boundary Layer Experiment) uses data from 12th of March 2020 to simulate mixed-phase clouds in the polar marine boundary layer region \cite{ccpp-scm-cases-v7}. 

\begin{figure}[htb]
    \centering
    {\includegraphics[width=\columnwidth]{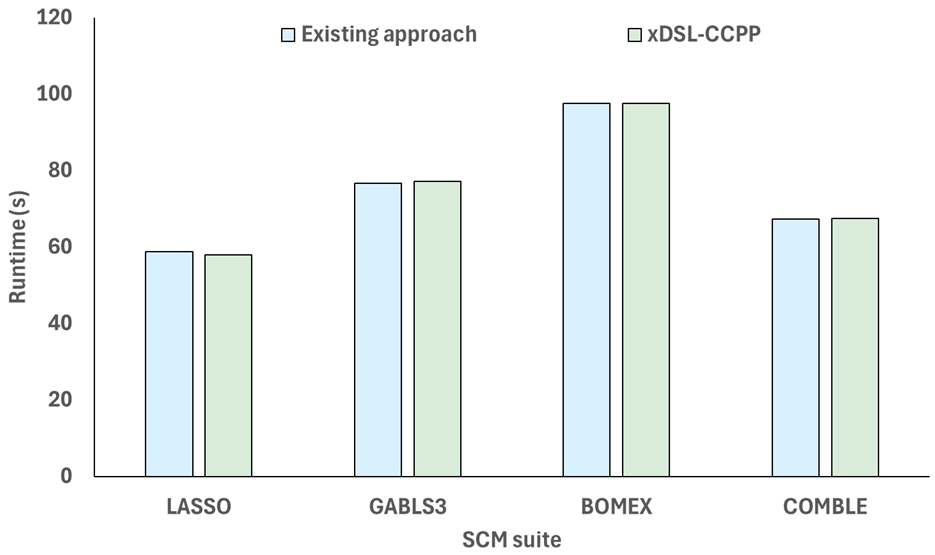}}
    \caption{Existing code generation against our approach for four SCM suites on Cirrus}
    \label{fig:scm}
\end{figure}

All runs simulate 1000 seconds, with a timestep of 0.1 seconds and reported results averaged over three runs. Figure \ref{fig:scm} illustrates performance of these SCM suites when built with the existing code generation tool compared to xDSL-CCPP, and it can be seen that, for each of these suites, performance is comparable between the two approaches. This confirms that xDSL-CCPP can be swapped in without incurring runtime penalty.

The runtime differences between the existing tool and xDSL-CCPP (Fortran) have a standard deviation of below 4\% across three runs. This is expected, because both approaches generate functionally equivalent interface code and runtime is dominated by the physics schemes, which are unchanged. The purpose of these experiments was therefore to confirm that xDSL-CCPP introduces no runtime penalty, and its benefits lie in maintainability and usability rather than boosting performance.

\section{Conclusions and further work}
\label{sec:conc}

In this paper we have explored a new Domain Specific Language (DSL), xDSL-CCPP, for configuring the execution of the Common Community Physics Package (CCPP). As HPC codes are modernised, a crucial question is how users can most appropriately drive them in a flexible manner, and the complexity traditionally associated with developing DSLs can be off-putting. We have demonstrated that modern compiler infrastructure provides a sea change and significantly reduces this overhead. Indeed, by formalising application configuration and interaction as a DSL, one is able to both reduce bespoke code complexity and improve the user interface compared to configuration files, which often grow organically over time. 

Whilst xDSL-CCPP increased compilation time compared to the current tool, the amount of bespoke code reduced by three times and the overall model runtime is unaffected. Moreover, our approach provides additional flexibility such as being able to infer metadata from Fortran source code and hence avoid the need for manually written, and maintained, metadata files. Furthermore, we demonstrated that by working with Intermediate Representation (IR) at all stages in the pipeline, it is trivial to develop new frontends and provide a choice around the user interface. This enabled a Python-based frontend to be developed, which addresses some of the inflexibility associated with XML, especially when expressing control flow. The reduction in bespoke code benefits tool maintainers, metadata inference removes a manual burden from scheme developers, and the Python frontend gives suite integrators a more expressive configuration interface.

For future work, we intend to explore automatic porting of suites to GPUs. There is a rich GPU ecosystem provided by MLIR, and previous work \cite{brown2024fully} demonstrated that existing MLIR passes were able to automatically transform serial loops in the \emph{scf} dialect into the \emph{gpu} dialect. 

\section*{Acknowledgements}
This work was funded by the CONTINENTS EPSRC project grant number EP/Z531170/1. For the purposes of open access, the author has applied a CC BY public copyright license to any Author Accepted Manuscript version arising from this submission.

\bibliographystyle{splncs04}
\bibliography{references}

\end{document}